%% file: maindocument.tex
\documentclass[manuscript,nonacm]{acmart}
\AtBeginDocument{%
  \providecommand\BibTeX{{%
    \normalfont B\kern-0.5em{\scshape i\kern-0.25em b}\kern-0.8em\TeX}}}

\setcopyright{CC}

\usepackage{subcaption}

\begin{document}

\input{0_title_page}
\input{1_abstract}

\input{2_key_words}
\input{3_introduction}
\input{4_motivation}
\input{5_methodology}

\input{6_results}
\input{7_discussion}
\input{8_conclusion}
\input{9_acknowledgments}
\input{10_references}

\end{document}

%% file: 0_title_page.tex
\title{Introducing Vibration for use in Interaction Designs to support Human Performance: A Pilot Study}

\author{Alexander D. Bincalar}
\affiliation{%
  \institution{University of Southampton}
  \city{Southampton}
  \country{UK}
}

\author{m.c. schraefel}
\affiliation{%
  \institution{University of Southampton}
  \city{Southampton}
  \country{UK}
}

\author{Christopher T. Freeman}
\affiliation{%
  \institution{University of Southampton}
  \city{Southampton}
  \country{UK}
}


%% file: 1_abstract.tex
\begin{abstract}
While vibration is a well-used output signal in HCI as part of haptic interaction, vibration outside HCI is used in many other modes to support human performance, from rehabilitation to cognition. In this late breaking work, we present preliminary positive results of a novel protocol that informs how vibration might be used to enrich HCI interventions for aspects of both health and intellectual performance. We also present a novel apparatus specifically designed to help HCI researchers explore different vibration amplitudes and frequencies for such applications.
  
\end{abstract}

%% file: 2_key_words.tex


\maketitle

%% file: 3_introduction.tex
\section{Introduction}
Vibration is a form of physical interaction that is perceived through mechanoreceptors \cite{iheanacho_2022} - specialised nerves that sense mechanical deformation (changes in joint angle, skin stretch, tendon shape change) and may be interpreted as contact or touch. Depending on the source, vibration can be felt very locally, on the skin (cutaneous, like a breeze on the skin), or in the whole body (kinesthetic, like feeling an earthquake as a tremor).

In HCI the use of vibration as an interaction paradigm has been mainly situated as a component of haptic interaction. Via smartphone or watch, we commonly experience vibration as a localised physical signal, where various kinds of vibrations are associated with different signals: a phone call, a timer ending, a cue to go left or right, an appointment, whether a virtual button has been successfully pressed, which just names a few of the many applications of haptic signalling. More recently, haptic interaction is also being explored for simulating touch and contact with virtual barriers in VR \cite{martin_2022} and AR \cite{bermejo_2021}.

Outside the interactive contexts of vibration for signaling or for simulation of contact, vibration has numerous direct applications. For example, in sports science and rehabilitation, two established areas where vibration in particular are \emph{Localised Vibration} and \emph{Whole Body Vibration}. For localised vibration, there are various forms of stimulation from devices that apply vigorous force to areas particularly around the lungs to loosen mucous in cystic fibrosis patients \cite{bradley_2010}, to massage guns that likewise vibrate an area of the body as a passive physiological rehabilitation pathway \cite{garcia-sillero_2021}. There are also targeted electrical stimulation approaches that create a sense of vibration. The electro-magnetic stimulation induces muscular activation by triggering muscle fibers to contract and then release that contraction at  varying rates \cite{seetohul_2021}. As vibration-sensitive nerves are embedded in muscle fibers, these contractions are experienced as vibration. This vibratory stimulation is used to induce effects from mild analgesia to passively triggering hypertrophy (muscle growth) \cite{rambhia_2022}. Whole body vibration is often produced by standing on a plate that moves at a particular frequency and amplitude \cite{cardinale_2005}. This approach has been used to accelerate fracture healing, blood circulation and pulmonary function, as well as to mitigate the effects of numerous neurological diseases \cite{orozi_2020}, not least vestibular/balance issues \cite{yilmaz_2019, Ardic_2021}.

While these vibration approaches are designed mainly for rehabilitation or strength building, because of the perception of vibration via our vestibular and proprioceptive systems \cite{yilmaz_2019}, recent research has been exploring how vibration can provide assistance to improve balance as well. For example, neck muscle vibration has been shown to have a positive impact on balance, gait speed and postural control \cite{wannaprom_2018, jamal_2020}. Low intensity ankle vibration has been shown to reduce the risk of falling in members of elderly populations by improving balance performance \cite{toosizadeh_2018}. With the same aim of reducing fall rates in the elderly, haptic/vibratory feedback systems applied to the soles of the feet have been shown to adjust users' centre of pressure \cite{tanaka_2009}. 

Based on these studies and work in our lab, we also hypothesise that vibratory stimulation that benefits balance would also benefit cognitive performance. A simple neurological equation for this proposition is that: the body is constantly sensing and testing our physiological state. Neuroscientist Joseph LeDoux characterised this sensing as threat/no threat detection \cite{ledoux_1998}. Therefore, the more stable/balanced the body, the lower the perceived threat, the more capacity for thinking.  

In this paper, we present the state of our current investigations into developing and testing a passive assistive interaction approach to enhance balance and cognitive executive functioning. In the following sections, we present: an overview of balance as a neurocognitive mechanism and how balance is assessed; we also review the theory of how vibration interacts with balance and cognition. From here we present our apparatus and pilot study to begin to explore the interplay of vibration, balance and cognitive performance. 

The results of the following pilot study suggest our approach offers a rich seam for future novel interaction techniques for human performance. The three main contributions from this work are, first, to present new applications of vibration for HCI to explore; second more specifically, to present vibration as a novel passive interaction pathway for benefit to both balance and cognitive performance, and third to present a novel mechanism to enable control of frequency and amplitude of vibration for future HCI studies.

%% file: 4_motivation.tex
\section{Context: Balance and Vibration}

The act of balancing is often an overlooked skill, despite it being essential to daily life. The time required to acquire a balance skill can range from months to years, and like any other skill, the ability to balance can decline due to inactivity \cite{cunningham_2020}. Inactivity means a lack of stimulating those parts of the nervous system that enable balance. For example sitting taxes our equilibrium far less than standing or walking. Furthermore, balance quality has also been shown to be an indicator of cognitive health and life expectancy \cite{tabara_2014, araujo_2022}. Therefore, looking at ways to interface with balance such that it can be maintained or improved is beneficial for multiple areas of health and well-being. Integrating balance better into work/life offers a novel opportunity space for HCI.  

The information used to maintain balance primarily comes from three sensory inputs: vision, the vestibular system, and the proprioceptors. Vision informs the person about their environment and how they are positioned within it \cite{bednarczuk_2021}, the vestibular system tracks the orientation and acceleration of the head (which is also used in conjunction with vision to stabilise gaze) \cite{khan_2013}, while proprioceptors keep track of the body’s self-locomotion (what limb is where moving how fast relative to other parts) \cite{gilman_2002}.

Through these sensory inputs, numerous ways to interface with balance, including vibration, have been explored. These include using virtual reality to create a more effective means of training balance compared with conventional training \cite{sarasso_2022}, applying electrical stimulation to the vestibular system via surface electrodes at the back of the ears (Galvanic Vestibular Stimulation \cite{sluydts_2020}) and using mechanical vibration on various locations of the body, such as the feet and ankles, with the primary aim of stimulating the proprioceptors \cite{tanaka_2009, toosizadeh_2018} and the neck for stimulating the vestibular system through the mechanoreceptors \cite{wannaprom_2018}. When using vibration, there are two main mechanoreceptors of interest, the Meissner corpuscles and the Pacinian corpuscles, which are sensitive to 30Hz and 250Hz frequencies, respectively \cite{iheanacho_2022}. Due to this physiological link, the effect these two frequencies have on balance and cognition will be investigated in this paper.

\subsection {The interplay of mechanisms between balance and vibration}

As previously mentioned, numerous studies have investigated the use of vibration for improving balance \cite{wannaprom_2018, tanaka_2009, toosizadeh_2018}, but this is usually in the context of improving balance in the elderly. Whole Body Vibration has also been investigated for this application \cite{orr_2015}, but it has also been researched for rehabilitating musculoskeletal problems in athletes \cite{sierra-guzman_2018}. From the above studies, there are two main approaches for influencing balance, that is to interface with proprioception and/or the vestibular system via stimulating the mechanoreceptors, which in turn, can stimulate the vestibular or muscles through neighbouring nerves and pathways.

\subsection {The interplay of vibration and cognition}

As well as being effective for treating musculoskeletal problems and improving blood circulation, among other problems \cite{orozi_2020}, Whole Body Vibration has also been shown to have a positive impact on cognition, improving attention, inhibition, memory and learning \cite{regterschot_2014, Freitas_2022}. However, there has been a lack of research into the area of how localised vibration can influence cognition, which is a feature of our work. If localised vibration can have a positive impact on cognition, it would be much more practical to implement that Whole Body Vibration, especially within a HCI context.

%% file: 5_methodology.tex
\section{Materials and Methods}
The goal of this study is to test three questions: (i) does vibration applied along a particular neuro-muscular pathway improve balance, (ii) is the effect improved using the particular frequencies that the nerves are sensitive to rather that a random frequency, and (iii) does the vibration also have a positive effect on cognitive performance? 

The following sections report on our pilot study, which is designed both to test our study protocol, and to get a sense of preliminary insights into these questions.

\subsection{Pilot Study \& Preliminary Exploration Results} 
To evaluate the effect of local passive vibration on balance and cognition, we developed a small device, described below, where we can control the frequency and amplitude of the vibration applied to the area.

\begin{figure}[h]
  \centering
  \includegraphics[width=0.45\linewidth]{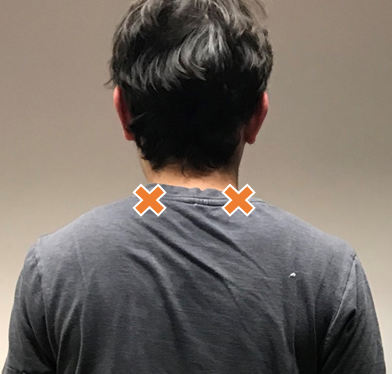}
  \caption{The orange crosses mark the two locations where vibration stimulation was applied}
  \Description{The two locations where the audio exciters were placed, which was on the back of the body, around the nape of the neck}
  \label{fig:vibration_location}
\end{figure}

The apparatus currently places the vibration on the trapezius muscle area: that is the area on either side of the upper spine (below cervical vertebrae 7 and above thoracic vertebrae T3, see Fig. \ref{fig:vibration_location}). 

One reason for choosing this location as a first test site for local vibration stimulation. Several studies in vibration and balance have focused on the neck area. Intriguingly these papers do not rationalise physiologically why they target these areas, rather than for example the lower limb that other studies also use for balance and vibration. 

Our rationale beyond following in these studies' footsteps, is several-fold. First, while balance is influenced by proprioception, vision and vestibular system combined, in this study we wish to focus our intervention on the vestibular component in particular. Feet and ankle stability are more associated with proprioceptive properties of balance \cite{robbins_1995} so we are moving effectively as far away from that area as we can for local stimulation.  

Second, a role of the vestibular system is to manage head position - not least for support of vision (the two work together for balance via the vestibular-occular reflex) \cite{khan_2013}. Head position is largely supported by the many muscles of the neck. The trapezius is the largest muscle supporting the head, connecting as well to the shoulders and mid back. These three sections each send signals to and from the brain via the same nerve: cranial nerve XI, the accessory nerve.  

Third, the accessory nerve runs along a similar pathway in the spine and to the limbs as does cranial nerve 8 - the vestibulocochlear nerve. CN8 synapses (connects) directly with nerves in C7 and C8 and into the thoracic (mid back) spinal nerves including the accessory, CN11. Thus information to manage movement of the upper limbs/back/neck connects with the vestibular system.  Our apparatus thus triggers these regions of the \textit{medial vestibulospinal tract}.

Of the many balance tests available, we chose
the Sharpened Romberg \footnote{\href{https://www.physio-pedia.com/index.php?title=Romberg_Test&oldid=319741}{Romberg Test}}.
\begin{comment}
\begin{figure}[ht]
  \centering
  \includegraphics[width=0.4\linewidth]{sections/figures/sharpened_romberg.png}
  \caption{The Sharpened Romberg stance.}
  \Description{A person standing in a Sharpened Romberg stance, which is a tandem stance with eyes closed}
  \label{fig:balance_tasks}
\end{figure}
\end{comment}
The Sharpened Romberg balance task is conducted without shoes, as shoes limit the joint range of motion, affecting proprioception, and shoe cushioning limits mechanoreceptive stimulation of the foot, all to different degrees depending on the stiffness/cushioning of the shoe. Thus excluding footwear gave us a more similar stance baseline.

Participants are asked whether they suffer from any vestibular lesions or other conditions that affect balance, whether they have any current musculoskeletal injuries and if they are hypersensitive to vibration. If a participant fits into any of these categories, then they are excluded from the study.

The advantage of the Sharpened Romberg, and with the position of the device as explained above, is that we can have some measurable control on the three factors of balance: eyes are closed, reducing visual cues; stance proprioception is in the same condition across participants, and our intervention is non-invasively triggering a vestibular pathway.   
 
\subsection{Cognitive Tests}
For cognition, the Word-Colour Interference test (Stroop test) \cite{stroop_1935} was used. The Stroop test measures a range of executive functions, from cognitive flexibility to attention, by testing how quickly someone can distinguish between the name of a colour written in text and the colour of that text. This test output multiple metrics, such as the average response time it takes to answer each trial correctly and incorrectly, and the difference between those two averages (which is known as the Stroop effect). Improvement in cognitive performance, specifically the Stroop test, has been observed in a Whole Body Vibration study, which could be due to the increase in participants' attention \cite{regterschot_2014}.

\subsection{Participants}
Our pilot study included 6 male participants (mean age 26, standard deviation 5.48). Of course for a larger study we balance the demographic. Our intent here has been to test our protocol. Participants were then randomly assigned to the Control or Intervention Groups. The study was approved by the University of Southampton Ethics committee, with ERGO number 79141.

The intervention group was run as a within group design where frequency of vibration was the main variable of interest. We therefore alternated the test of frequencies – 30/250Hz, 250/30Hz to minimise any ordering effect on performance. 

\subsection{Protocol}
\begin{figure}[ht]
  \centering
  \includegraphics[width=0.6\linewidth]{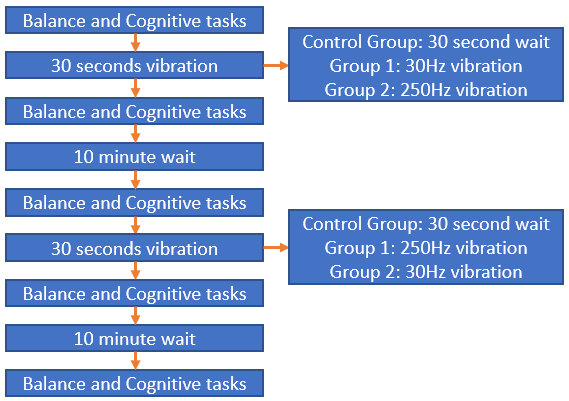}
  \caption{A diagram of the protocol, displaying each step. Group 1 and 2 are the intervention groups that experience experiences in the order of 30/250Hz and 250/30Hz, respectively}
  \Description{A diagram of the protocol that lists out each task, intervention and rest period in chronological order}
  \label{fig:protocol}
\end{figure}

Figure \ref{fig:protocol} describes the experimental protocol.

In the protocol, INTERVENTION participants perform a pre-intervention cognitive assessment and balance assessment. They then have a vibration stimulus applied; they repeat the balance and cognitive assessment tasks. Participants sit for ten minutes, then they redo the protocols using the alternative frequency; they do another round of balance and cognitive assessments, another ten minute sit, and then a final balance and cognitive assessment.  

Based on Wannaprom's recent study investigating the affect of neck muscle vibration on balance \cite{wannaprom_2018}, we are likewise applying vibration for 30 seconds. In future work we will also look at time/dose effects.

\subsection{Apparatus}
\subsubsection{Generating Vibration:}
Previous studies have generally used eccentric rotating mass (ERM) vibration motors or Linear Resonant Actuators (LRAs) to apply vibration. As this study is exploring multiple frequencies, an audio exciter was selected as the vibration source as amplitude, frequency and motion of vibration can all be individually controlled electronically. For other vibration sources such as ERM motors, frequency and amplitude are dependent on each other, while LRAs are tuned to a specific vibration frequency (based on the resonant frequency of the design) \footnote{\label{footnote:vibration_motors}\href{https://www.precisionmicrodrives.com/motors/vibration-motors}{Information on Vibration Motors}}. Audio exciters, allow these parameters to be independently controlled, giving great freedom on how to apply and change the stimulus. However, this comes at the cost of complexity, as an audio exciter requires a custom driver to control its vibration (LRAs is also similar in that regard), whilst ERM vibration motors only require a DC signal to operate.

The vibration is applied through two 25mm audio exciters \footnote{\href{https://www.daytonaudio.com/product/1175/daex25ct-4-coin-type-25mm-exciter-10w-4-ohm}{DAEX25CT-4 Coin Type 25mm Exciter 10W 4 Ohm}}, which are rated up to 10W, but only 3W of power was applied through them. As Audio exciters enable any waveform to pass through them, a custom sine wave generator was designed with a user configurable amplitude and frequency to control the audio exciters, such that the appropriate power levels and frequencies could be set for this experiment, but also so that these parameters could be reconfigured for future studies.

\begin{figure}[h]
  \centering
  \includegraphics[width=0.9\linewidth]{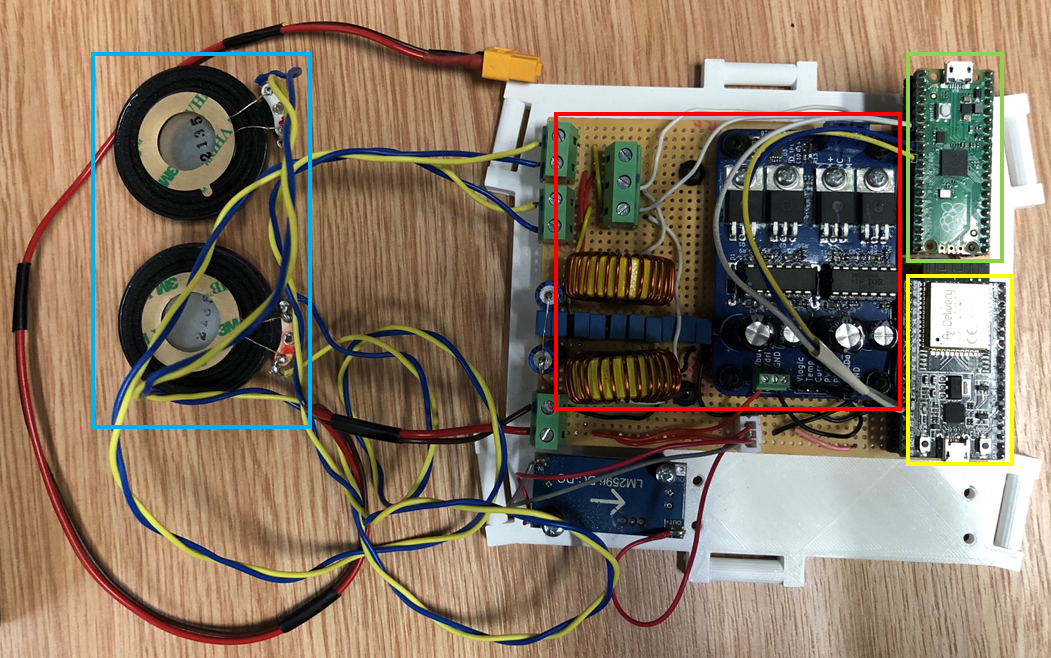}
  \caption{A diagram of the hardware, showing the audio exciters within the light blue border, an H-bridge and low pass filter inside the red border, an RP2040 microcontroller within the green border and an ESP32 microcontroller inside the yellow border.}
  \label{fig:hardware}
\end{figure}

\begin{figure}[ht]
  \centering
  \includegraphics[width=1\linewidth]{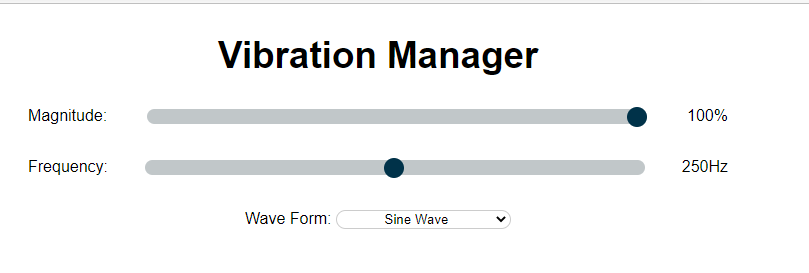}
  \caption{The web application used to configure the frequency, magnitude and wave form for the audio exciters.}
  \label{fig:web_app}
\end{figure}

Figure \ref{fig:hardware} shows the hardware developed for this research and Figure \ref{fig:web_app} shows the web application that communicates the desired frequencies and magnitudes to the hardware. An ESP32 hosts the web server, allowing any device to connect to it if on the same local network. The ESP32 then communicates the selected magnitude and frequency information to the RP2040, which acts as a waveform generator by using Pulse Width Modulation (PWM). The outputted pulses from the RP2040 are amplified by an H-bridge, connected to a low pass filter which averages the pulses into a smooth waveform (this system forms a class-D amplifier). The waveform then travels to the audio exciters which vibrate to the selected frequency and magnitude. The shape of this waveform can be controlled by varying the duty cycles of the PWM from the RP2040, allowing any form of wave to be produced (including audio), but for the study sine waves were used as they only have a first harmonic frequency.

\subsubsection{Measuring Balance:}
Balance can be measured in a number of ways. Common measures in clinical practice and research include:
the participant’s balance time, their change in postural sway and their change in rate of postural sway.
In these assessments, balance time includes how long the participant can hold the balance position before having to change position. For example in a Sharpened Romberg's test that has eyes closed and one foot right in front of the other, time would be stopped when the participant had to move a foot to stay upright. 
Postural sway is the amount of movement the participant makes during the balance task \cite{alsubaie_2019}. Better balance results in a finer ability to keep the body in the same position, resulting in more accurate and smaller movements for maintaining balance.

We used the Microsoft Azure Kinect both for calculating balance time and degree of postural sway, as it has been proven to work well for this application \cite{antico_2021}.  The Azure Kinect was set up to record and save depth recordings of the balance tasks, as depth is used by the Body Tracking SDK to calculate joint positions \footnote{\href{https://learn.microsoft.com/en-us/azure/kinect-dk/body-sdk-download}{Microsoft Azure Kinect Body Tracking SDK}}. Using this SDK, we developed a custom program to track the naval spine position (as this point is close to the centre of mass of a person when standing). The change in position of the X coordinate (the sideways motion of the participant) was calculated by taking the maximum and minimum positions during the balance task. This result was normalised with respect to the pre-intervention balance test to track the improvement as a percentage change.

%% file: 6_results.tex
\section{Results}

\begin{figure}[ht]
    \centering
    \begin{subfigure}{0.49\textwidth}
        \includegraphics[width=\textwidth]{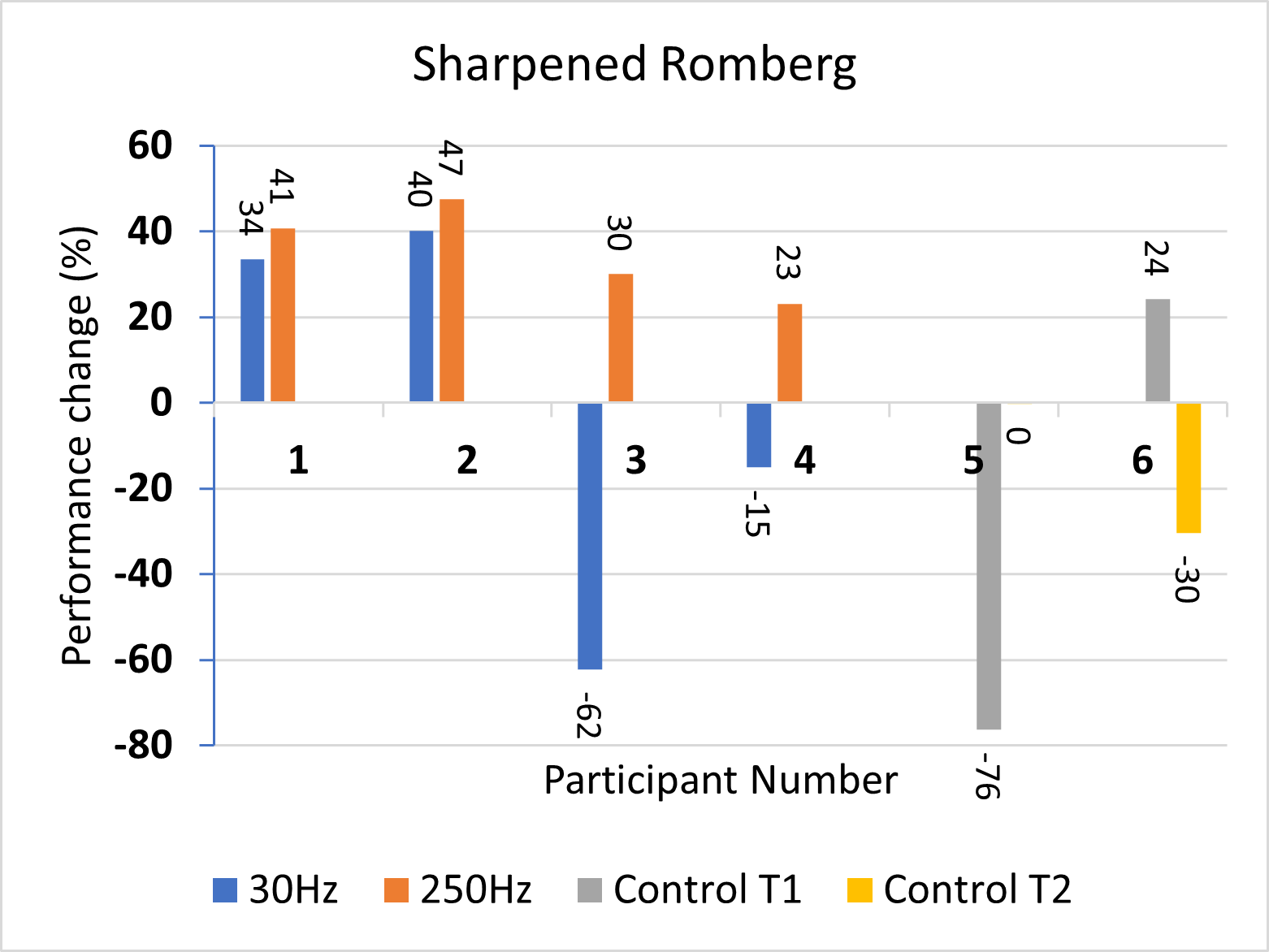}
        \caption{Sharpened Romberg Results}
        \label{fig:sharpened_romberg_results}
    \end{subfigure}
    \begin{subfigure}{0.49\textwidth}
        \includegraphics[width=\textwidth]{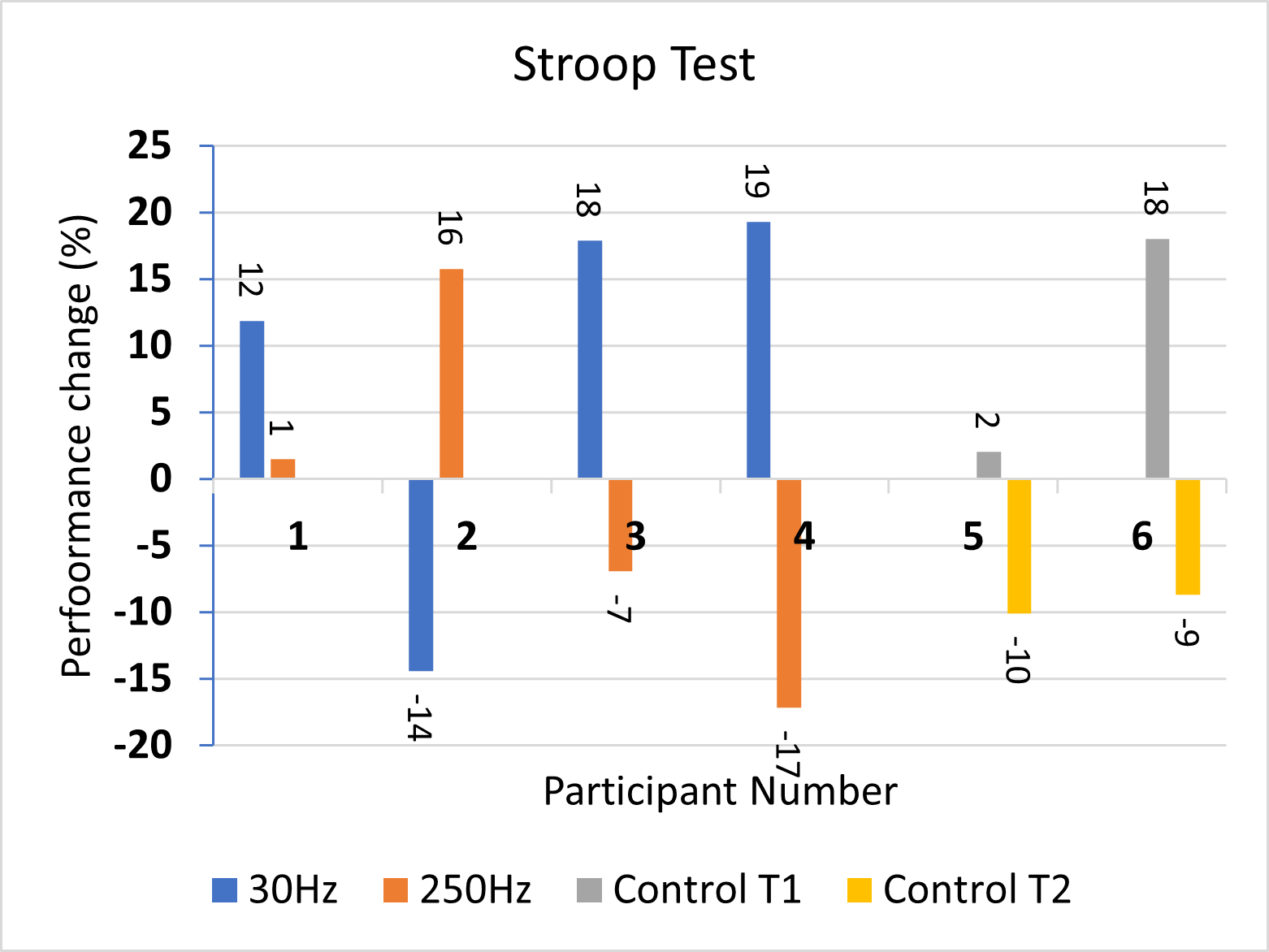}
        \caption{Stroop Test Results}
        \label{fig:stroop_test_results}
    \end{subfigure}
    \caption{Results for the Sharpened Romberg and the Stroop Test. As the control group repeated part of the experiment with no intervention, there are two results from each participant in that group. Control T1 is the first set of results and control T2 is the second set.}
    \label{fig:results}
\end{figure}

Figure. \ref{fig:results} displays the results from the sharpened Romberg and Stroop test. As some participants had significant differences in their absolute results, all the results have been normalised based on the pre-intervention results, which acts as a baseline. The results are expressed as a percentage that indicates much of a performance increase/decrease has been gained post-intervention.

\textbf{Balance:} The results indicate that 250Hz vibration shows a positive impact on the sharpened Romberg, as all 4 participants had a reduction in their postural sway during the balancing task, ranging from 23\% to a 47\% improvement in performance (\ref{fig:sharpened_romberg_results}). 

\textbf{Cognitive Executitve Functioning:} Three out of four participants showed an improvement in their Stroop test congruence with the 30Hz vibration  (see \ref{fig:stroop_test_results}). With the sample size it's not possible to say if the non-responder is an outlier or part of a pattern. Again a larger cohort will help unpack this result. Still the result is suggestive, as the results converge towards one direction. 

Most interestingly, both frequencies appear to have different applications, with 250Hz correlating with benefits to balance while 30Hz had an affect on cognition. Overall, in this pilot study, there are no ordering effects. 

%% file: 7_discussion.tex
\section{Discussion and Future Work}
A goal of the work presented has been to explore the potential benefit of developing interactive computational devices to deliver vibratory stimulation to support better balance and cognition. The approach would complement and enlarge the space for use of vibration beyond haptics. 

The results of the pilot study suggest that there is a positive effect on both balance and cognition, sufficiently worthwhile to carry on to a full study. We hope these preliminary results encourage others to engage with this area.

\subsection{Limitations}
A key limit of a pilot study is of course its size: we cannot draw statistical results from what is effectively an assessment of a protocol. Likewise our sample is too homogeneous to make any inference about population effects. Similarly we have only used one amplitude, and one period of vibration dose and these may be factors around intensity of effect or its duration. 

\subsection{Future Work}
The results of this pilot study suggest that the protocol is sound and that there is an effect. 

An interesting result that we did not anticipate is that different frequencies appear to have different applications, with 30Hz appearing to have more cognitive benefit, while 250Hz has a greater impact on balance. One possibility for this response is due to the difference in the speed of signalling from the specific mechanoreceptors that these two different frequencies target. We look forward to exploring this further.

Most of our participants were very active. We are keen to investigation the effect across more diverse demographics. Likewise we will be looking at factors like sedentariness and foot wear trends as possible factors influencing effects.

Fundamentally, for HCI work, we, and we hope other researchers, will use this data to explore applications. For example, if a knowledge work experiences fatigue, could they lean up against a vibration device, and dial in time/frequency to help them recover focus - and see how their attention or task performance improves? 

%% file: 8_conclusion.tex
\section{Conclusion}

In this paper, we offer the following contributions:

1. \textit{VIBRATION AS EFFECT:} We introduce less explored ways in HCI to use haptics/vibration for interaction. In particular these include interaction design opportunities to incorporate vibration to support human performance. 

2. \textit{APPARATUS:} We present a novel device to enable specific control of key parameters - frequency and amplitude of vibration - to be adjusted to explore dose/response effects.

3. \textit{Balance-Cognition Connection:} We present the protocol and pilot study assessment of a novel interaction approach for HCI that may enhance multiple aspects of human performance, operationalising (i) vestibular-perturbation for its affects on (ii) balance and (iii) cognition.

While from the preliminary results we see from the pilot study, there is potential for interactive technology to use vibration for balance and cognitive benefits. Our intent in offering these preliminary findings is to encourage other researchers to explore this area along with us. 

%% file: 9_acknowledgments.tex
%%\begin{acks}

%%\end{acks}

%% file: 10_references.tex
\bibliographystyle{ACM-Reference-Format}
\bibliography{bibliography}